\documentclass[aps,prb,twocolumn,showpacs,amsmath,amssymb]{revtex4}
\usepackage{graphicx}
\usepackage{dcolumn}
\usepackage{bm}

\begin{document}
\title{Weak ferromagnetism in antiferromagnets: Fe$_{2}$O$_{3}$ and La$_{2}$CuO$_{4}$}

\author{V. V. Mazurenko$^{1,2}$ and V. I. Anisimov$^{1,2}$}
\affiliation{$^{1}$Theoretical Physics and Applied Mathematics Department, Urals State Technical University, Mira Street 19,  620002
Ekaterinburg, Russia \\
$^{2}$Institute of Metal Physics, Russian Academy of Sciences, 620219 Ekaterinburg GSP-170, Russia}%

\date{\today}

\begin{abstract}
The problem of weak ferromagnetism in antiferromagnets due to
canting of magnetic moments was treated using Green's function
technique. At first the eigenvalues and eigenfunctions of the
electronic Hamiltonian corresponding to collinear magnetic
configuration are calculated which are then used to determine
first and second variations of the total energy as a function of
the magnetic moments canting angle. Spin-orbit coupling is taken
into account via perturbation theory. The results of calculations
are used to determine an effective spin Hamiltonian.
This Hamiltonian can be mapped on conventional
spin Hamiltonian that allows to determine parameters of isotropic
and anisotropic (Dzyaloshinskii-Moriya) exchange interactions. The
method was applied to the typical antiferromagnets with weak
ferromagnetism Fe$_{2}$O$_{3}$ and La$_{2}$CuO$_{4}$. The obtained
directions and values of the magnetic moments canting angles are
in a reasonable agreement with experimental data.
\end{abstract}

\maketitle

Heisenberg Hamiltonian is a basis of most theoretical
investigations of transition metal compounds magnetism.
\cite{Heisenberg,Anderson,dagotto} The essential part of these
investigation is determination of exchange interaction parameters
J$_{ij}$. It can be done in a phenomenological way by fitting
those parameters to reproduce experimental data (temperature
dependence of magnetic susceptibility and magnon dispersion curves
obtained in inelastic neutron scattering measurements).\cite{VOPO}
However much more physically appealing is to obtain them in
{\it{ab-initio}} calculations. In most cases it was done via
calculating total energy values for various different magnetic
moments configurations. Mapping on Heisenberg Hamiltonian gave a
system of linear equation for J$_{ij}$ (for example see
Ref.~\onlinecite{Pickett}). This procedure becomes inconvenient
for the systems with a large number of long range competing
exchange interactions like in (VO)$_{2}$P$_{2}$O$_{7}$,
NaV$_{2}$O$_{5}$, Cu$_{2}$Te$_{2}$O$_{5}$X$_{2}$ (X=Br,Cl), {\it
etc}. \cite{lemmens}

In 1987 A.I. Lichtenstein {\it{et al.}} \cite{Liechtenstein}
proposed the calculation method that does not use total energy
differences. They determined  exchange interaction parameters via
calculating  second variation of total energy  $\delta^2 E$ for
small deviation of magnetic moments from the collinear magnetic
configuration. The expression for this second variation was
derived analytically and required for its evaluation calculation
of the integral over production of one-electron Green functions.
This method was since then successfully applied to the various
transition metal compounds.
\cite{solovyev1,Korotin99,Korotin00,Mn12}

The combination of low symmetry and spin-orbit coupling was shown
by Dzyaloshinskii \cite{dzialoshinski} and Moriya \cite{moriya} to
give rise to anisotropic exchange coupling. Moriya has shown how
the processes involving  an additional virtual transition due to
spin-orbit coupling can cause an anisotropic exchange interaction
as a correction to the isotropic Anderson superexchange term and
introduced new term in spin Hamiltonian which is
Dzyaloshinskii-Moriya interaction (DM). Solovyev {\it{et al.}}
\cite{solovyev} had shown that Dzyaloshinskii-Moriya interaction
parameters can be calculated using perturbation theory and Green's
function technique and described the canting of magnetic moments
of LaMnO$_{3}$. Recently, Katsnelson and Lichtenstein
\cite{katsnelson} derived the general expression for
Dzyaloshinskii-Moriya interaction term in LDA++ approach.

This paper is devoted to the problem of first-principles
theoretical description of weak ferromagnetism in
antiferromagnets, specifically to the task of calculating  weak
ferromagnetic moment value and direction of spin canting. For this
we consider first and second variations of total energy of the
system at small deviation of magnetic moments from collinear
configuration with spin-orbit coupling introduced as a
perturbation using Green's function technique. We show that there
is an additional on-site term that was not taken into account in
previous work, \cite{solovyev} which gives significant
contribution to weak ferromagnetic moment. Basing on the results
of the calculations we propose effective single site Hamiltonian.
This Hamiltonian is sufficient
for solving the problem of spin canting but it also can be
rewritten to the conventional form containing isotropic and
anisotropic exchange interaction terms. We have applied our method
for weak ferromagnetism in Fe$_{2}$O$_{3}$, the classical system
which was used by Moriya in his pioneering work, \cite{moriya} and
in antiferromagnetic cuprate La$_{2}$CuO$_{4}$ in low temperature
orthorhombic phase, estimated ferromagnetic moments values on the
metallic ions in these compounds and determined the plane of spin
canting.

Briefly, this paper is organized as follows. In Sec.I we describe
the method for calculation spin Hamiltonian parameters responsible
for magnetic moments canting. Sec.II contains the
results of our calculations for Fe$_{2}$O$_{3}$ and
La$_{2}$CuO$_{4}$ crystals. In Sec.III we discuss and briefly
summarize our results.

\section{Method}
\newcommand{\ud}{\mathrm{d}}
According to the Andersen's ``local force theorem''
\cite{machintosh,methfessel,heine} the total energy variation
$\delta E$ under the small perturbation from the ground state
coincides with the sum of one-particle energy changes for the
occupied states at the fixed ground state potential. In the first
order for the perturbations of the charge and spin densities one
can find the following relation \cite{Liechtenstein}:
\begin{eqnarray}
\delta E = \int_{-\infty}^{E_{F}} \ud \epsilon \, \epsilon \, \delta n(\epsilon) = \epsilon_{F} \, \delta z
- \int_{-\infty}^{E_{F}} \ud \epsilon \, \delta N (\epsilon)
\nonumber \\
= - \int_{-\infty}^{E_{F}} \, \ud \epsilon \, \delta N (\epsilon),
\end{eqnarray}
here $n(\epsilon)=\ud N/ \ud \epsilon$ is the density of electron state, N($\epsilon$) is integrated density of electron state and
$\epsilon_{F}$ is Fermi energy. In the case of magnetic
excitation the change of total number of electrons $\delta z$ equals zero.
The Green function G is formally expressed in the usual way $G=(\epsilon-H)^{-1}$.
One can express density of states and integrated density of states via Green function G:
\begin{eqnarray}
n(\epsilon)= -\frac{1}{\pi} \, Im \, Sp \,G(\epsilon)
\end{eqnarray}
and
\begin{eqnarray}
N(\epsilon) = -\frac{1}{\pi} \, Im \, Sp \, \ln(\epsilon-H).
\end{eqnarray}
Then the variation of integrated density of state is given by
\begin{eqnarray}
\delta N(\epsilon)=\frac{1}{\pi} \, Im \, Sp \, [\delta H \, G].
\end{eqnarray}
Therefore the first and second variations of total energy of the system take the following forms:
\begin{eqnarray}
\delta E=  - \frac{1}{\pi} \, \int_{-\infty}^{E_{F}} \ud \epsilon \, Im \, Sp \,  (\delta H \, G) \label{firstvarE}
\end{eqnarray}
and
\begin{eqnarray}
\delta^{2} E=  - \frac{1}{\pi} \, \int_{-\infty}^{E_{F}} \ud \epsilon \, Im \, Sp \,  (\delta^{2} H \, G \, + \,
\delta H \, G \, \delta H \, G).
\end{eqnarray}

Operator of spin rotation on the site $j$ on the angle $|\vec
\delta \phi|$ around direction $\vec{n} \, = \, \frac{\vec {\delta
\phi}} {|\vec {\delta \phi}|}$ is given by
\begin{eqnarray}
\widehat U=e^{\frac{1}{2}\, i \, \vec{\delta \phi} \, \widehat {\vec{ \sigma}}}
\end{eqnarray}
where $\widehat {\vec \sigma} \, = \, (\widehat \sigma_{x}, \widehat \sigma_{y}, \widehat \sigma_{z})$ are Pauli matrices.
For small $|\vec {\delta \phi}|$ values we can expand the spin rotation operator in following way
\begin{eqnarray}
\widehat U \, = \, 1 \, + \, \frac{1}{2}\, i \, \vec{\delta \phi} \, \widehat {\vec{ \sigma}} -
\frac{1}{8} \, (\vec{\delta \phi} \, \widehat {\vec{ \sigma}})^{2}.
\end{eqnarray}
New Hamiltonian of the system after rotation of the spin on j site around direction $\vec{n}$ on the angle $|\vec {\delta \phi}|$
\begin{eqnarray}
\widehat{\widetilde{H}}= \widehat{U^{\dag}} \widehat{H}
\widehat{U}
\end{eqnarray}
The first variation over the angle of rotation is expressed in
following form:
\begin{eqnarray}
\delta \widehat H= \frac{1}{2} \, i \, \vec{\delta \phi} \, [\widehat{H}, \widehat{\vec{\sigma}}] \label{operfirstvar}
\end{eqnarray}

In the basis $|ilm\sigma\rangle$ (where $i$ is site, $l$ is
orbital quantum number, $m$ is magnetic quantum number and
$\sigma$ is spin index) the Hamiltonian matrix takes the form
$H_{ilm,jlm'}^{\sigma \sigma'}  =  \langle ilm \sigma| \widehat{H}
|jlm' \sigma' \rangle$. For simplicity below we drop the index of
orbital and magnetic quantum numbers and leave spin and site
indexes. We assume that without spin-orbit interaction the ground
state corresponds to the collinear magnetic configuration at which
all spin moments lie along z axis. Therefore the Hamiltonian
matrix H$_{ij}^{\sigma \sigma'}$ is diagonal in the spin subspace
\begin{displaymath}
H_{ij}=
\left( \begin{array}{cc}
{\bf H_{ij}^{\uparrow}} & 0 \\
0 & {\bf H_{ij}^{\downarrow}}
\end{array} \right) .
\end{displaymath}
One can rewrite the first variation of Hamiltonian Eq.(\ref{operfirstvar}) in the following form
\begin{eqnarray}
\delta H_{jj} \, = \, i \, \delta \phi^{x}_{j} \left( \begin{array}{cc}
0 & \frac{\Delta_{j}}{2}   \\
-\frac{\Delta_{j}}{2}  & 0
\end{array} \right) \,
+\, \delta \phi^{y}_{j} \left( \begin{array}{cc}
0 & \frac{\Delta_{j}}{2} \\
\frac{\Delta_{j}}{2}  & 0
\end{array} \right), \label{firstdervH}
\end{eqnarray}
where $\Delta_{j} \, = \, {\bf H^{\uparrow}_{jj}} \, - \, {\bf H^{\downarrow}_{jj}}$.
It is easy to show that the second variation of Hamiltonian is given by
\begin{eqnarray}
\delta^{2} H_{jj} = \delta^{2} \phi^{x}_{j} \left( \begin{array}{cc}
- \frac{\Delta_{j}}{2} & 0 \\
0 & \frac{\Delta_{j}}{2}
\end{array} \right)  +  \delta^{2} \phi^{y}_{j} \left( \begin{array}{cc}
- \frac{\Delta_{j}}{2} & 0 \\
0 & \frac{\Delta_{j}}{2}
\end{array} \right). \label{seconddervH}
\end{eqnarray}
The rotation of spin moment around z axis doesn't change the energy of the system therefore the term with $\delta \phi^{z}$ is absent in
Eq.(\ref{firstdervH},\ref{seconddervH}).

Then we take into account the spin-orbit coupling via perturbation theory. The Green function
in the first order perturbation theory with respect to the spin-orbit coupling can be written as
\begin{eqnarray}
\tilde G_{ij} \, = \, G_{ij} \, + \, \sum_{k} G_{ik} \, H^{so}_{k} \, G_{kj},
\end{eqnarray}
here $H^{so}_{k}=\lambda_{k} \, \vec{L} \, \vec{S}$, i, j and k
denote site, G$^{ij}$ is Green function of system with collinear
magnetic configuration, and $\lambda_{k}$ is spin-orbit coupling
constant. The first variation of total energy Eq.(\ref{firstvarE})
takes the form:
\begin{eqnarray}
\delta E \, = \, - \frac{1}{\pi} \, \sum_{i} \int_{-\infty}^{E_{F}} \ud \epsilon \, Im \, Sp
\nonumber \\
\times (\delta H_{i} \, G_{ii} \, + \, \sum_{k} \, \delta H_{i} \, G_{ik} \, H^{so}_{k} \, G_{ki} ) \label{firstvar}.
\end{eqnarray}
The first term in Eq.(\ref{firstvar}) is zero. The second term can be expressed as a following sum
\begin{eqnarray}
\delta E \, = \, \sum_{i} A^{x}_{i} \, \delta \phi^{x}_{i} \, + \, A^{y}_{i} \, \delta \phi^{y}_{i}, \label{firstvariat}
\end{eqnarray}
where
\begin{eqnarray}
A^{x}_{i} =  \sum_{k} B^{x}_{ik} =
\, - \frac{1}{2 \pi}  \int_{-\infty}^{E_{F}} \, \ud \epsilon  \, Re
\nonumber \\
\times \sum_{k} \, (\Delta_{i} \,
G_{ik}^{\downarrow} \, H^{so}_{k \, \downarrow \uparrow} \,
G_{ki}^{\uparrow}
- \, \Delta_{i} \, G_{ik}^{\uparrow} \, H^{so}_{k \, \uparrow \downarrow} \, G_{ki}^{\downarrow})
\end{eqnarray}
and
\begin{eqnarray}
A^{y}_{i}  =  \sum_{k} B^{y}_{ik}  =
\, - \frac{1}{2 \pi} \int_{-\infty}^{E_{F}} \, \ud \epsilon  \, Im
\nonumber \\
\times \sum_{k} \, (\Delta_{i} \, G_{ik}^{\downarrow}
\, H^{so}_{k \, \downarrow \uparrow} \, G_{ki}^{\uparrow} \,
+ \, \Delta_{i} \, G_{ik}^{\uparrow} \, H^{so}_{k \, \uparrow \downarrow} \, G_{ki}^{\downarrow}).
\end{eqnarray}
We consider the situation when all spins lie along z axis and
therefore the rotation around it does not change the energy of
system. In order to find A$^{z}_{i}$ component of the magnetic
torque vector $\vec A_{i}$  we change the coordinate system in the
following way (x,y,z)$\rightarrow$(z,y,-x) (rotation around y
axis):
\begin{eqnarray}
\tilde H^{so} \, =  \,
\frac{1}{2} \left( \begin{array}{cc}
1 & 1 \\
-1 & 1
\end{array} \right) \, \left( \begin{array}{cc}
H^{so}_{\uparrow \uparrow} & H^{so}_{\uparrow \downarrow} \\
H^{so}_{\downarrow \uparrow} & H^{so}_{\downarrow \downarrow}
\end{array} \right) \, \left( \begin{array}{cc}
1 & -1 \\
1 & 1
\end{array} \right). \label{Hso}
\end{eqnarray}
Therefore A$^{x}_{i}$ component in new
coordinate system is A$^{z}_{i}$ in the old one:
\begin{eqnarray}
\label{Az}
 A^{z}_{i}  =  \sum_{ik} B^{z}_{ik}   =
\, - \frac{1}{4\pi}   \int_{-\infty}^{E_{F}}  \, \ud \epsilon  \, Re
\nonumber \\
\times \sum_{k} \, (\Delta_{i} \, G_{ik}^{\uparrow} \,
(H^{so}_{k \, \uparrow \uparrow} \, - \, H^{so}_{k \, \downarrow \downarrow}) \, G_{ki}^{\downarrow}
\nonumber \\
- \Delta_{i} \, G_{ik}^{\downarrow} \,
(H^{so}_{k \, \uparrow \uparrow} \, - \, H^{so}_{k \, \downarrow \downarrow}) \, G_{ki}^{\uparrow}).
\end{eqnarray}

In contrast to the first variation $\delta E$, the second
variation of total energy $\delta^{2} E$ for small deviations of
magnetic moments from ground-state collinear magnetic
configuration has nonzero value without taking into account
spin-orbit coupling:
\begin{eqnarray}
\delta^{2} E \, = \, - \frac{1}{\pi} \, \int_{-\infty}^{E_{F}} \ud \epsilon \, Im \, Sp \, (\frac{1}{2} \, \sum_{i} \delta^{2}H_{ii}G_{ii}
\nonumber \\
+ \, \frac{1}{2} \, \sum_{j} \, \delta^{2}H_{jj}G_{jj} \, + \, \sum_{ij} \, \delta H_{i} \, G_{ij} \, \delta H_{j} \, G_{ji}), \label{secondvar}
\end{eqnarray}
where
\begin{eqnarray}
Sp \, (\delta^{2} H_{ii} \, G_{ii}) \, = \, \frac{1}{2} \delta^{2} \phi^{x}_{i} \Delta_{i} (G_{ii}^{\downarrow}  \, -  \, G_{ii}^{\uparrow})
\nonumber \\
+ \, \frac{1}{2} \delta^{2} \phi^{y}_{i} \Delta_{i} (G_{ii}^{\downarrow} \, -  \, G_{ii}^{\uparrow})
\end{eqnarray}
and
\begin{eqnarray}
Sp \, (\delta H_{i} \, G_{ij} \, \delta H_{j} \, G_{ji}) \, = \, \frac{1}{2} \, \delta \phi^{x}_{i} \, \delta \phi^{x}_{j} \,
(\Delta_{i} \, G^{\downarrow}_{ij} \, \Delta_{j} \, G^{\uparrow}_{ji})
\nonumber \\
+ \, \frac{1}{2} \, \delta \phi^{y}_{i} \, \delta \phi^{y}_{j} \, (\Delta_{i} \, G^{\downarrow}_{ij} \, \Delta_{j} \, G^{\uparrow}_{ji}).
\end{eqnarray}
Using the condition $G_{ii}^{\uparrow} \, - \, G_{ii}^{\downarrow} \, = \, (G^{\uparrow}
\Delta \, G^{\downarrow})_{ii} \, =\, \sum_{j} G_{ij}^{\uparrow} \Delta_{j} \, G_{ji}^{\downarrow}$ one may rewrite Eq.(\ref{secondvar}) in following form:
\begin{eqnarray}
\delta^{2} E \, = \, \frac{1}{4\pi} \, \int_{-\infty}^{E_{F}} \ud \epsilon \, Im
\, \sum_{ij} (\Delta_{i} \,
G_{ij}^{\downarrow} \, \Delta_{j} \, G_{ji}^{\uparrow})
\nonumber \\
\times ((\delta \phi_{i}^{x} \, - \delta \phi_{j}^{x})^{2} \, +
 \, (\delta \phi_{i}^{y} \, - \delta \phi_{j}^{y} )^{2}). \label{exchange}
\end{eqnarray}
One can see that the Eq.(\ref{exchange}) contains only $x$ and $y$
components of $\vec{\delta \phi}$. In order to include $z$
component one can use the same rotation of coordinate system as
for the site magnetic torque vector $\vec A$ Eq.(\ref{Hso}). Finally,
we obtain the following function of the total energy over angle
$\vec {\delta \phi}$:
\begin{eqnarray}
\Delta E \, = \, \sum_{i} \vec A_{i} \, \vec{\delta \phi_{i}} \, +
\, \frac{1}{2} \, \sum_{ij} \, J_{ij} \, |\vec{\delta \phi_{i}} \, - \, \vec{\delta \phi_{j}}|^{2}, \label{totalexpan}
\end{eqnarray}
where
\begin{eqnarray}
J_{ij} \, = \, \frac{1}{4\pi} \, \int_{-\infty}^{E_{F}} \ud \epsilon \, Im (\Delta_{i} \,
G_{ij}^{\downarrow} \, \Delta_{j} \, G_{ji}^{\uparrow}).
\end{eqnarray}

The aim of this paper is description of canted magnetism in
transition metal compounds caused by spin-orbit coupling. For this
we have used the expression Eq.(\ref{totalexpan}) for the total
energy as a function of canting angle. In order to solve the
problem of the weak ferromagnetism in antiferromagnets we suppose
that the crystal is an antiferromagnet containing two sublattices
$1$ and $2$, with the same canting angle for the atoms belonging
to the same sublattice.
With this assumption  the Eq.(\ref{totalexpan}) is reduced to the
following form:
\begin{eqnarray}
\Delta E \, = \, \vec A_{1} \, \vec{\delta \phi_{1}} \, + \, \vec A_{2} \, \vec{\delta \phi_{2}} \, + \,
\sum_{j>1} J_{1j} \, |\vec{\delta \phi_{1}} \, - \, \vec{\delta \phi_{2}}|^{2}.
\end{eqnarray}
Our results for Fe$_{2}$O$_{3}$ and La$_{2}$CuO$_{4}$ demonstrated
that $\vec A_{1}$ = - $\vec A_{2}$ (torque vector has an opposite
sign for the atoms belonging to the different sublattices). That
gives:
\begin{eqnarray}
\Delta E \, = \, \vec A_{1} \, (\vec{\delta \phi_{1}} - \vec{\delta \phi_{2}}) \, + \,
 \sum_{j>1} J_{1j}  \, |\vec{\delta \phi_{1}} - \vec{\delta \phi_{2}}|^{2}. \label{functional}
\end{eqnarray}
If we further suppose that the deviations of magnetic moments from
the average direction defined by $\vec{\delta \phi_{i}}$ have the
same absolute value but different sign for both sublattices, then
Eq.(\ref{functional}) takes the following form (we suppose that
magnetic moments lie in plane perpendicular to site magnetic
torque vector $\vec A$ and canting occurs in the same plane)
\begin{eqnarray}
\Delta E  \, = \, 2 \, \vec A_{1} \, \vec{\delta \phi_{1}} \, + \,
4 \, \sum_{j>1} J_{1j} \, |\vec{\delta \phi_{1}}|^{2}.
\end{eqnarray}
Then we find the value of $|\vec{\delta \phi_{1}}|$ where $\Delta E$ has a minimum:
\begin{eqnarray}
\label{canting}
 |\vec{\delta \phi_{1}}| \, = \, \frac{|\vec A_{1}|}{4 \, \sum_{j>1} J_{1j}}.
\end{eqnarray}

The next step is to establish a connection
between Eq.(\ref{totalexpan}) and conventional spin Hamiltonian
\begin{eqnarray}
H \, =  \,H_{DM} +H_{exch} \, = \, \sum_{i \not= j} \vec D_{ij} \, [\vec e_{i} \times \vec e_{j}]
\nonumber \\
+ \, \sum_{i \not= j} \, J_{ij} \, \vec e_{i} \vec e_{j}, \label{spinHam}
\end{eqnarray}
where e$_{i}$ is unit vector in the direction of the $i$th site
magnetization, J$_{ij}$ is exchange interaction and $\vec D_{ij}$
is Dzyaloshinskii-Moriya vector. One can rewrite the second term
in Eq.(\ref{spinHam}) as $H_{exch} \, = \, \sum_{ij} J_{ij} |\vec
e_{i}| \, |\vec e_{j}| cos (\theta_{ij})$. In the limit of small
canting angle values we can assume that $cos(\theta_{ij}) \, = \,
1-|\vec{\delta \phi_{i}} - \vec{\delta \phi_{j}}|^2/2$ and
exchange interaction energy for antiferromagnetic configuration
has a form:
\begin{eqnarray}
\Delta H_{exch} \, = \, \frac{1}{2} \sum_{ij} J_{ij} \, |\vec{\delta
\phi_{i}} - \vec{\delta \phi_{j}}|^2.
\end{eqnarray}
Therefore in the limit of small $\vec {\delta \phi}$ we can
directly map the second term of total energy variation
Eq.(\ref{totalexpan}) onto first term in spin Hamiltonian
Eq.(\ref{spinHam}).

The first term in Eq.(\ref{totalexpan}) describes the deviation of
the spin moment on the site $i$ from the initial collinear spin
configuration direction. We assume that this initial spin
direction on the site $i$ is defined by the direction of Weiss
mean-field $\vec H^{WF}_{i} \, = \, \sum_{j (\not = i)} J_{ij} \,
\vec e_{j}$ (corresponding unit vector is $\vec e^{\, \, 0}_{i}=\frac{\vec
H^{WF}_{i}}{|\vec H^{WF}_{i}|}$). Therefore we can map the first
term in Eq.(\ref{totalexpan}) on the spin Hamiltonian
\begin{eqnarray}
H_{dev} \, = \, \sum_{i} \vec A_{i} \, [\vec e^{\, \, 0}_{i} \times  \vec e_{i}] \label{dev}
\end{eqnarray}
describing the deviation of spin moments away from the direction
$\vec e^{\, \, 0}_{i}$ of external Weiss field. (We have used here the
connection between rotation vector $\vec{\delta \phi_{i}}$ and the
change of the magnetic moment unit vector $\delta\vec e_{i}=\vec
e_{i}-\vec e^{\, \, 0}_{i}$: $\delta\vec e_{i}=[\vec{\delta
\phi_{i}}\times \vec e^{\, \, 0}_{i}]$.) In order to demonstrate the
connection between $\vec A_{i}$ and $\vec D_{ij}$ one can rewrite
the Eq.(\ref{dev}) in the following form:
\begin{eqnarray}
H_{dev} \, = \, \frac{1}{2} (\sum_{i} \vec A_{i} \, [\frac{\vec H^{WF}_{i}}{|\vec H^{WF}_{i}|} \times  \vec e_{i}]
\nonumber \\
+ \, \sum_{j} \vec A_{j} \, [\frac{\vec H^{WF}_{j}}{|\vec H^{WF}_{j}|} \times  \vec e_{j}]).
\end{eqnarray}
Using our definition of $\vec H^{WF}_{i}$ we obtain:
\begin{eqnarray}
H_{dev} \, = \, \frac{1}{2} (\sum_{ij} \frac {\vec A_{i}}{|\vec H^{WF}_{i}|} J_{ij} [\vec e_{j} \times  \vec e_{i}]
\nonumber \\
+ \, \sum_{ij} \frac{\vec A_{j}}{|\vec H^{WF}_{j}|} J_{ij} [\vec e_{i} \times  \vec e_{j}]).
\end{eqnarray}
This gives us the following expression for parameter $\vec
D_{ij}$ of spin Hamiltonian Eq.(\ref{spinHam}):
\begin{eqnarray}
\vec D_{ij} \, = \, \frac{1}{2} \, J_{ij} \,(\frac{\vec A_{j}}{|\vec H^{WF}_{j}|} \, - \, \frac{\vec A_{i}}{|\vec H^{WF}_{i}|}).
\end{eqnarray}
Therefore the components of Dzyaloshinskii-Moriya interaction vector are given by
\begin{eqnarray}
D^{x}_{ij} \, = \, - \frac{1}{4\pi} \,J_{ij} \, \int_{-\infty}^{E_{F}} \ud \epsilon \, Re \, \sum_{k}
\nonumber \\
\times ( \frac{\Delta_{j} \, G_{jk}^{\downarrow} \,
H^{so}_{k \, \downarrow \uparrow} \, G_{kj}^{\uparrow} \, -
\, \Delta_{j} G_{jk}^{\uparrow} \, H^{so}_{k \, \uparrow \downarrow} \, G_{kj}^{\downarrow}}{|\vec H^{WF}_{j}|}
\nonumber \\
- \, \frac{\Delta_{i} \, G_{ik}^{\downarrow} \, H^{so}_{k \, \downarrow \uparrow} \, G_{ki}^{\uparrow} \, -
\, \Delta_{i} G_{ik}^{\uparrow} \, H^{so}_{k \, \uparrow \downarrow} \, G_{ki}^{\downarrow}}{|\vec H^{WF}_{i}|}),
\end{eqnarray}

\begin{eqnarray}
D^{y}_{ij} \, = \, - \frac{1}{4\pi} \, J_{ij} \, \int_{-\infty}^{E_{F}} \ud \epsilon \, Im \, \sum_{k}
\nonumber \\
\times (\frac{\Delta_{j} \, G_{jk}^{\downarrow}
\, H^{so}_{k \, \downarrow \uparrow} \, G_{kj}^{\uparrow} \, +
\, \Delta_{j} G_{jk}^{\uparrow} \, H^{so}_{k \, \uparrow \downarrow} \, G_{kj}^{\downarrow}}{|\vec H^{WF}_{j}|}
\nonumber \\
- \, \frac{\Delta_{i} \, G_{ik}^{\downarrow} \, H^{so}_{k \, \downarrow \uparrow} \, G_{ki}^{\uparrow} \, +
\, \Delta_{i} G_{ik}^{\uparrow} \, H^{so}_{k \, \uparrow \downarrow} \, G_{ki}^{\downarrow}}{|\vec H^{WF}_{i}|}),
\end{eqnarray}

\begin{eqnarray}
D^{z}_{ij} \, = - \, \frac{1}{8\pi} \, J_{ij} \, \int_{-\infty}^{E_{F}} \ud \epsilon \, Re \, \sum_{k}
\nonumber \\
\times (\frac{\Delta_{j} \, G_{jk}^{\uparrow} \,
(H^{so}_{k \, \uparrow \uparrow} \, - \, H^{so}_{k \, \downarrow \downarrow}) \, G_{kj}^{\downarrow}}{|\vec H^{WF}_{j}|}
\nonumber \\
- \frac{\Delta_{j} \, G_{jk}^{\downarrow} \,
(H^{so}_{k \, \uparrow \uparrow} \, - \, H^{so}_{k \, \downarrow \downarrow}) \, G_{kj}^{\uparrow}}{|\vec H^{WF}_{j}|}
\nonumber \\
-\frac{\, \Delta_{i} \, G_{ik}^{\uparrow} \,
(H^{so}_{k \, \uparrow \uparrow} \, - \, H^{so}_{k \, \downarrow \downarrow}) \, G_{ki}^{\downarrow}}{|\vec H^{WF}_{i}|}
\nonumber \\
+ \frac{\Delta_{i} \, G_{ik}^{\downarrow} \,
(H^{so}_{k \, \uparrow \uparrow} \, - \, H^{so}_{k \, \downarrow \downarrow}) \, G_{ki}^{\uparrow}}{|\vec H^{WF}_{i}|}).
\end{eqnarray}

We have obtained more general expression for Dzyaloshinskii-Moriya
interaction parameter in comparison with those presented in paper.
\cite{solovyev} There are two kind of contributions into magnetic
torque vector $\vec A_{i}$: on-site interaction $\vec B_{ii}$
(absent in work \cite{solovyev}) and intersite interaction $\vec
B_{ik}$(i $\not =$ k). We have found that on-site contribution in
magnetic torque $\vec A$ which was not considered before plays an
important role in weak ferromagnetism  description.

We have applied the calculation scheme developed above to the
typical antiferromagnets with weak ferromagnetism Fe$_{2}$O$_{3}$
and La$_{2}$CuO$_{4}$ in low-temperature orthorhombic phase. In
order to calculate Green functions corresponding to the collinear
spin configurations we used LDA+U approach \cite{Anisimov}
realized in LMTO method within Atomic Sphere Approximation.
\cite{Andersen}

\section{results}
\subsection{Fe$_{2}$O$_{3}$}

Weak ferromagnetism or weak non-collinearity of essentially antiparallel magnetic moments was first observed in $\alpha$-hematite,
$\alpha$-Fe$_{2}$O$_{3}$.\cite{smith} The trigonal crystal of Fe$_{2}$O$_{3}$ has R${\bar 3}$c space group. Depending on temperature
$\alpha$-hematite may occur in two different antiferromagnetic states: at T $<$ 250 K the spins are along the trigonal axis, and at
250 K $<$ T $<$ 950 K they lie in one of the vertical planes of symmetry making a small angle $1.1 \times 10^{-3}$ with basal
plane.\cite{Flanders,Bodker} In the latter case the
$\alpha$-Fe$_{2}$O$_{3}$ has a net ferromagnetic moment. Dzyaloshinskii has shown that the spin superstructure gives rise to a nonvanishing
antisymmetric spin coupling vector which is parallel to the trigonal axis. Moriya \cite{moriya} gave
phenomenological Dzyaloshinski's explanation a microscopic
footing by means of Anderson's perturbation approach to magnetic superexchange.

Sandratskii $et$ $al$. \cite{sandratskii} have performed the
calculation based on the local approximation to spin-density
functional theory (LSDA) using the fully relativistic version of
ASW method. In spite of the well-known problem that the LSDA has
with a proper determination of the energy gap in semiconducting
and insulating materials, the authors \cite{sandratskii} succeeded
in describing a weak ferromagnetism and obtained
the ferromagnetic moment of about 0.002 $\mu_{B}$. In the
present study we treat the problem of description of weak
ferromagnetism in Fe$_{2}$O$_{3}$ using perturbation theory.

\begin{figure}[!h]
\centering
\includegraphics[width=0.35\textwidth]{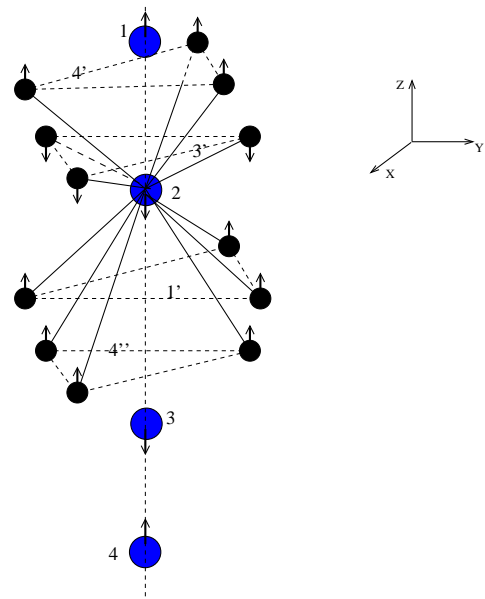}
\caption {Crystal structure of $\alpha$-Fe$_{2}$O$_{3}$. Large circles are Fe atoms which belong the smallest unit cell used in the LDA+U calculations.
The small circles are Fe atoms which surround Atom 2. Arrows denote the magnetic configuration used in our calculation.}
\end{figure}

Electronic structure of $\alpha$-hematite calculated using the
standard LDA+U approximation \cite{Anisimov} with on-site Coulomb
interaction parameters U = 5 eV, J = 0.88 eV and structure data
from \cite{newnham} is in a good agreement with previous
theoretical calculations.\cite{Amrit} We have obtained the
magnetic moment of 4.1 $\mu_{B}$ per Fe atom. This value is a
little smaller than those obtained in experiments (4.6-4.9
$\mu_{B}$). The energy gap value of 1.67 eV is also slightly
underestimated comparing with experimental data (2.14 eV in
paper \cite{benjelloun}).

The Brillouin zone integration has been
performed in the grid generated by using (6;6;6) divisions. The
energy integration has been performed in the complex plane by
using 800 energy points on the rectangular contour. The calculated
isotropic exchange interactions and contributions in site magnetic torque $\vec
A_{2}$ are presented in Table I. The simplified crystal structure
and the interaction paths are shown in Fig.1.

One can see that the obtained interaction picture is more
complicated that those Moriya examined in order to describe the
weak ferromagnetism in $\alpha$-hematite.\cite{moriya} There are
strong isotropic exchange interaction with third and fourth
neighbours. This agrees well with experimental results
\cite{samuelsen} and theoretical predictions.\cite{goodenough}
The summary exchange of atom 2 with nearest neighbours is given by $J_{2}=\sum_{i \not = 2} J_{2 \,
i} \, = \,$ 189.26 meV.

\begin {table}
\centering
\caption [Bset]{Isotropic exchange interaction parameter and different contributions in components of site magnetic torque of Fe$_{2}$O$_{3}$ (in
meV). The couplings with negative sign are ferromagnetic.
d$_{ij}$ is distance between i and j atoms in a.u.
$\vec R_{ij}$ is radius vector from i site to j site in units of the lattice constant (5.49 a.u.).}
\label {basisset}
\begin {tabular}{lcccccc}
  \hline
  \hline
  (i, j)  &d$_{ij}$ & $\vec R_{ij}$ & J$_{ij}$  & B$^{x}_{ij}$ & B$^{y}_{ij}$ & B$^{z}_{ij}$ \\
  \hline
  (2,2)  & 0  & (0;0;0)                  & 0        &  0  &  0  &  0.162 \\
  \hline
  (2,1)  & 5.45  & (0;0;-0.99)         & 8.576   &  0       &  0      &  0.005  \\
  (2,3') & 5.60 & (-0.5;-0.86;0.20)    &  -7.3     &  -0.036  &  0.015  &  0.001  \\
  (2,3') & 5.60 & (1;0;0.20)            & -7.3     &  0.032   &  0.023  &  0.001  \\
  (2,3') & 5.60 & (-0.5;0.86;0.20)     &  -7.3     &  0.004   & -0.038  &  0.001  \\
  (2,1') & 6.36 & (0.5;-0.86;-0.58)   & 25.224  &  0.071   &  0.019  &  -0.14  \\
  (2,1') & 6.36 & (-1;0;-0.58)         & 25.224  &  -0.052  &  0.052  &  -0.14   \\
  (2,1') & 6.36 & (0.5; 0.86;-0.58)   & 25.224  &  -0.019  & -0.071  &  -0.14   \\
  (2,4') & 6.99 & (0.5;-0.86;0.79)    & 17.502  &  0.168   &  0.063  &  0.101   \\
  (2,4') & 6.99 & (-1;0;0.79)          & 17.502  & -0.139   &  0.114  &  0.101  \\
  (2,4') & 6.99 & (0.5;0.86;0.79)     & 17.502  & -0.029   & -0.178  &  0.101   \\
  (2,4'') & 6.99 & (-0.5;-0.86;-0.79) & 17.502  &  0.128   &  0.094  &  0.076  \\
  (2,4'') & 6.99 & (1;0;-0.79)         & 17.502  &  0.017   & -0.158  &  0.076  \\
  (2,4'') & 6.99 & (-0.5;0.86;-0.79)  & 17.502  & -0.145   &  0.064  &  0.076  \\
  \hline
  \hline
\end {tabular}
\end {table}

The components of site magnetic torque vector on the atom 2 are
given by $A^{x}_{2}= \sum_{i} B^{x}_{2i} \, = \,$ 0 eV,
$A^{y}_{2}= \sum_{i} B^{y}_{2i} \, = \, $ 0 eV and $A^{z}_{2}=
\sum_{i} B^{z}_{2i} \, = \, $ 0.281 meV. One can see that the
on-site interaction $B^{z}_{22}$ gives the main contribution in
$\vec A_{2}$. We have calculated also the site magnetic torque of
$\vec A_{1}$ which has the following components (0;0;-0.281), the
same value but the opposite sign comparing with $\vec A_{2}$. The
value of canting angle of $|\vec{\delta \phi}| \, = \, 0.4 \times
10^{-3}$ calculated with Eq.(\ref{canting}) is of the correct order of
magnitude but is more than two time smaller than experimental data
$1.1 \times 10^{-3}$ (Ref.~\onlinecite{Flanders,Bodker}). The reason for the
difference between experimentally observed and calculated here
values of canting angle could be the necessity to take into
consideration the higher order terms with respect to spin-orbit
coupling which were not considered in the present study.

It is easy to show that in the case when all spin lie along z axis
there is no canting of the spin moments. On the other hand if
direction of Weiss field is perpendicular to z axis the canting
exists and the system has weak ferromagnetic moment. This picture
fully agrees with experimental and theoretical data. \cite{moriya,Flanders,Bodker}

\subsection{La$_{2}$CuO$_{4}$}

In the case of the cuprates Dzyaloshinskii-Moriya interaction
is the leading source of anisotropy, since single-ion anisotropy
does not occur due to the S=$\frac{1}{2}$ nature  of the spins on
the Cu$^{2+}$ sites. The experimental data \cite{Thio,Kastner}
demonstrate that in case of low temperature orthorhombic phase the
spins do not lie exactly in the Cu-O planes, but are canted out of
the plane by a small angle (0.17$^{\circ}$). Coffey and coworkers
\cite{Coffey} made complete examination of the anisotropic
exchange interaction in orthorhombic phase based on a symmetry
consideration. They assumed rotation axis of the the CuO$_{6}$ as
a direction of antisymmetric exchange interaction and
obtained that the spins are canted in plane which is perpendicular
Dzyaloshinskii-Moriya vector.

\begin{figure}[!h]
\centering
\includegraphics[width=0.3\textwidth]{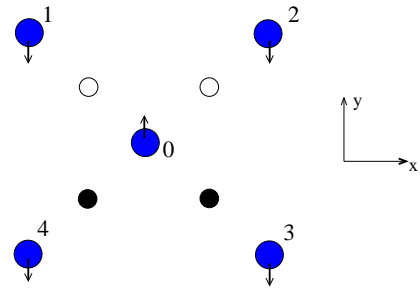}
\caption {The schematic crystal structure of La$_{2}$CuO$_{4}$ in low temperature orthorhombic phase. The open circles
are oxygen atoms which are tilted up out of the Cu plane, the black circles are oxygen atoms tilted down out the Cu plane.
The big gray circles are copper atoms. Arrows denote the magnetic configuration used in LDA+U
calculations with spin moments lie along z axis.}
\end{figure}

The first attempt at a microscopic calculation of the
Dzyaloshinskii-Moriya anisotropy for La$_{2}$CuO$_{4}$ in low
temperature orthorhombic and tetragonal phases was made by Coffey,
Rice, and Zhang.\cite{Rice} They have neglected the symmetric
anisotropic exchange interaction which is of the second order with
respect to the spin-orbit coupling and can be written in spin
Hamiltonian in the following form:
\begin{eqnarray}
H_{ij} \, =  \, \vec{S_{i}} \, M_{ij} \,\vec{S_{j}} \nonumber,
\end{eqnarray}
where M$_{ij}$ is a symmetric $3 \, \times \, 3$ tensor.

Shekhtman, Entin-Wolhman, and Aharony \cite{shekhtman} have shown
that symmetric anisotropic exchange interaction can not be
neglected, since its contribution to the magnetic energy is of the
same order of magnitude as that of the antisymmetric anisotropic
exchange interaction (Dzyaloshinskii-Moriya). On the basis of our
results (as we show below) we can conclude also that taking into
account the second order terms with respect to the spin-orbit
coupling must be important.

We have performed the LDA+U calculations for La$_{2}$CuO$_{4}$ in
low temperature orthorhombic phase using structural data for Nd
doped La$_{2}$CuO$_{4}$,\cite{Axe} with on-site Coulomb
interaction parameters U = 10 eV, J =1 eV (the same as used in work
\cite{Korotin}). The  schematic structure of Cu-O layer of
La$_{2}$CuO$_{4}$ in low temperature orthorhombic phase is
presented in Fig.2.

The experimental value of the energy gap is reported to be about 2 eV (Ref.~\onlinecite{cooper}).
Our gap value of 1.94 eV is in a good agreement with experimental data.
The calculated magnetic moment value on Cu atom is 0.61 $\mu_{B}$ which also agrees well with
experiment.\cite{Yamada}

\begin {table}[!h]
\centering
\caption [Bset] {Isotropic exchange J$_{0 j}$ and
the components of different contributions in site magnetic torque $\vec A_{0}$ (in meV).
$\vec R_{ij}$ is radius vector from i site to j site in units of the lattice constant (10.14 a.u.).}

\label {basisset}
\begin {tabular}{lccccc}
  \hline
  \hline
  (i, j)  & $\vec R_{ij}$ & J$_{ij}$ \quad \quad  & B$^{x}_{ij}$ \quad \quad & B$^{y}_{ij}$ \quad \quad & B$^{z}_{ij}$ \\
  \hline
  (0,0)   & (0;0;0)    & 0  \quad \quad       &  0.101 \quad \quad &  0 \quad \quad &  0      \\
  \hline
  (0,1)   & (-0.5;0.5;0)  & 14.576 \quad \quad  &  0.020 \quad \quad &  -0.032 \quad \quad & -0.005   \\
  (0,2)   & (0.5;0.5;0)   & 14.576 \quad \quad  &  0.020 \quad \quad &   0.032 \quad \quad &  0.005  \\
  (0,3)   & (0.5;-0.5;0)  & 14.576 \quad \quad  &  0.020 \quad \quad &  -0.032 \quad \quad & -0.005  \\
  (0,4)   & (-0.5;-0.5;0) & 14.576 \quad \quad  &  0.020 \quad \quad &   0.032 \quad \quad &  0.005  \\
  \hline
  \hline
\end {tabular}
\end {table}

We have performed calculations of isotropic exchange
interactions and the different contributions to site magnetic
torque components (Table II) using the energy integration in the
complex plane with 700 energy points on the rectangular contour
and the Brillouin zone integration has been performed in the grid
generated by using (6;6;6) divisions. The obtained values of
exchange interaction parameters are in a good agreement with
results of previous calculations for low temperature tetragonal
phase \cite{Korotin} and experimental estimations.\cite{Thio} The
exchange interactions with next neighbours are negligibly small.
The summary exchange and the components of site magnetic torque
are given by $J_{0} \, = \sum_{i \not = 0} J_{0 \, i} \, = \,$
58.304 meV, $A^{x}_{0} \, = \,$ 0.18 meV, $A^{y}_{0} \, = \, $ 0
meV and $A^{z}_{0} \, = \, $ 0 eV. We obtained that $\vec
A_{2}$=(-0.18;0;0), again of the same value but the opposite sign
comparing with $\vec A_{0}$. It means that the system has net
ferromagnetic moment if spins lie in plane which is perpendicular
to $x$ axis, which is axis of rotation of oxygen octahedra. This
fully agrees with results of previous theoretical works.\cite{Thio,Coffey,Rice}
The obtained value of canting angle $|\vec{\delta \phi}| \, = \, 0.77
\times 10^{-3}$ is about three times smaller than those
experimentally observed $2.2 \div 2.9 \times 10^{-3}$
(Ref.~\onlinecite{Thio,Kastner}).

Again as for Fe$_{2}$O$_{3}$ we expect that the inclusion of second order spin-orbit coupling terms could improve the agreement. Such calculations are in
progress.

\section{Conclusion}
We present a method for calculation of spin Hamiltonian parameters responsible for magnetic moments canting. The
effective Hamiltonian for canted magnetism was proposed. We shown
that the parameters of this model Hamiltonian can be obtained from
first-principles calculations. Using the developed method we
described the weak ferromagnetism in Fe$_{2}$O$_{3}$ and
La$_{2}$CuO$_{4}$. It was shown that on-site contribution $\vec
B_{ii}$ in site magnetic torque $\vec A_{i}$ plays the crucial
role for net ferromagnetic moment of Fe$_{2}$O$_{3}$ and
La$_{2}$CuO$_{4}$ in low temperature orthorhombic phase.

\section{ACKNOWLEDGMENTS}
We would like to thank A.I. Lichtenstein who initiated this
investigation. We also wishes to thank F. Mila, M. Troyer, T.M. Rice, M. Sigrist, M. Elhajal 
and M.A. Korotin for helpful discussions. This work is
supported by the
scientific program "Russian Universities"  yp.01.01.059 and
Russian Foundation for Basic Research grant RFFI 04-02-16096.


\begin{thebibliography}{99}
\bibitem{Heisenberg}
W. Heisenberg,
Z. Physik {\bf 49,} 619 (1928).

\bibitem{Anderson}
P.W. Anderson,
Phys. Rev. {\bf 115,} 2 (1959);
Solid State Physics {\bf 14,} 99 (Academic, New York 1963).

\bibitem{dagotto}
E. Dagotto and T.M. Rice,
Science {\bf 271,} 618 (1996).

\bibitem{VOPO}
A.W. Garrett, S.E. Nagler, D.A. Tennant, B.C. Sales, and T. Barnes,
Phys. Rev. Lett. {\bf 79,} 745 (1997).

\bibitem{Pickett}
W.E. Pickett,
Phys. Rev. Lett. {\bf 79,} 1746 (1997).

\bibitem{lemmens} P. Lemmens, G. G\"untherodt, and C. Gros,
Physics Reports {\bf 375,} 1 (2003).

\bibitem{Liechtenstein}  A.I. Lichtenstein, M.I. Katsnelson, V.P. Antropov, and V.A. Gubanov,
J. Magn. Magn. Mater. {\bf 67,} 65 (1987).

\bibitem{solovyev1} I.V. Solovyev and K. Terakura,
Phys. Rev. B {\bf 58,} 15496 (1998).

\bibitem{Korotin99}
M.A. Korotin, I.S. Elfimov, V.I. Anisimov, M. Troyer, and D.I. Khomskii,
Phys. Rev. Lett. {\bf 83,} 1387 (1999).

\bibitem{Korotin00}
M.A. Korotin, V.I. Anisimov, T. Saha-Dasgupta, and I. Dasgupta,
J. Phys.: Condens. Matter {\bf 12,} 113 (2000).

\bibitem{Mn12}
D.W. Boukhvalov, A.I. Lichtenstein, V.V. Dobrovitski, M.I. Katsnelson, B.N. Harmon, V.V. Mazurenko, and V.I. Anisimov,
Phys. Rev. B {\bf 65,} 184435 (2002).

\bibitem{dzialoshinski}  I. Dzyaloshinskii,
J. Phys. Chem. Solids {\bf 4,} 241 (1958).

\bibitem{moriya}  Toru Moriya,
Phys. Rev. {\bf 120,} 91 (1960).

\bibitem{solovyev} I.V. Solovyev, N. Hamada and K. Terakura,
Phys. Rev. Lett. {\bf 76,} 4825 (1996).

\bibitem{katsnelson}  M.I. Katsnelson and A.I. Lichtenstein,
Phys. Rev. B {\bf 61,} 8906 (2000).

\bibitem{machintosh}  A.R. Machintosh and O.K. Andersen, in: Electrons at the Fermin Surface,
ed. M. Springford (Cambridge Univ. Press, London, 1980) p. 149.

\bibitem{methfessel}  M. Methfessel and J. Kubler,
J. Phys. F {\bf 12,}  141 (1982).

\bibitem{heine} V. Heine, in: Solid State Physics,
ed. H. Ehrenreich et al. (Academic Press, New York, 1980) p. 1.

\bibitem{Anisimov} V.I. Anisimov, J. Zaanen, and O.K. Andersen,
Phys. Rev. B {\bf 44,} 943 (1991);
V.I. Anisimov, F. Aryasetiawan, and A.I. Lichtenstein, J. Phys.: Condens Matter {\bf 9,} 767 (1997).

\bibitem{Andersen} O.K. Andersen and O. Jepsen,
Phys. Rev. Lett. {\bf 53,} 2571 (1984);
O.K. Andersen, Z. Pawlowska, and O. Jepsen,
Phys. Rev. B {\bf 34,} 5253 (1986).

\bibitem{smith}  T. Smith,
Phys. Rev. {\bf 8,} 721 (1916); L. Neel, Rev. Mod. Phys. {\bf 25,} 58 (1953).

\bibitem{Flanders} P.J. Flanders and J.P. Remeika,
Philos. Mag. {\bf 11,} 1271 (1965).

\bibitem{Bodker} F. Bodker, M.F. Hansen, C.B. Koch, K. Lefmann, and S. Morup,
Phys. Rev. B {\bf 61,} 6826 (2000).

\bibitem{sandratskii} L.M. Sandratskii, M. Uhl, and J. K\"ubler,
J. Phys.: Condensed Matter {\bf 8,} 983 (1996);
L.M. Sandratskii and J. K\"ubler,
Europhys. Lett. {\bf 33,} 447 (1996).

\bibitem{newnham} R.E. Newnham and Y.M. de Haan,
Z.Kristallogr {\bf 117,} 235 (1962).

\bibitem{Amrit} A. Bandyopadhyay, J. Velev, W.H. Butler, S.K. Sarker, and O. Bengone,
Phys. Rev. B {\bf 69,} 174429 (2004).

\bibitem{benjelloun} D. Benjelloun, J.-P. Bonnet, J.-P. Doumerc, J.-C. Launay, M. Onillon, and P. Hagenmuller,
Mater. Chem. Phys. {\bf 10,} 503 (1984).

\bibitem{samuelsen} E.J. Samuelsen and G. Shirane,
Phys. Stat. Sol. {\bf 42,} 241 (1970).

\bibitem{goodenough} J.B. Goodenough,
Phys. Rev. {\bf 117,} 1442 (1960).

\bibitem{Thio} T. Thio, T.R. Thurston, N.W. Preyer, P.J. Picone, M.A. Kastner, H.P. Jenssen, D.R. Gabbe, C.Y. Chen,
R.J. Birgeneau, and A. Aharony,
Phys. Rev. B {\bf 38,} 905 (1988).

\bibitem{Kastner} M.A. Kastner, R.J. Birgeneau, T.R. Thurston, P.J. Picone, H.P. Jenssen, D.R. Gabbe, M. Sato, K.
Fukuda, S. Shamoto, Y. Endoh, K. Yamada, and G. Shirane,
Phys. Rev. B {\bf 38,} 6636 (1988).

\bibitem{Coffey} D. Coffey, K.S. Bedell, and S.A. Trugman,
Phys. Rev. B {\bf 42,} 6509 (1990).

\bibitem{Rice} D. Coffey, T.M. Rice and F.C. Zhang,
Phys. Rev. B {\bf 44,} 10112 (1991).

\bibitem{shekhtman} L. Shekhtman, O. Entin-Wohlman, and A. Aharony,
Phys. Rev. Lett. {\bf 69,} 836 (1992).

\bibitem{Axe} J.D. Axe and M.K. Crawford,
J. Low Temp. Phys. {\bf 95,} 271 (1994).

\bibitem{Korotin} V.I. Anisimov, M.A. Korotin, I.A. Nekrasov, Z.V. Pchelkina, and S. Sorella,
Phys. Rev. B {\bf 66,} 100502 (2002).

\bibitem{cooper} S.L. Cooper, G.A. Thomas, A.J. Millis, P.E. Sulewski, J. Orenstein, D.H. Rapkine, S-W.
Cheong, and P.L. Trevor,
Phys. Rev. B {\bf 42,} 10785 (1990).

\bibitem{Yamada} K. Yamada, E. Kudo, Y. Endoh, Y. Hidaka, M. Oda, M. Suzuki and T. Murakami,
Solid State Commun. {\bf 64,} 753 (1987).

\end{thebibliography}
\end{document}